# Small $x$ Contributions to the Structure Function $F_L(x, Q^2)$

Johannes Blümlein

*DESY–Institut für Hochenergiephysik, Zeuthen,
Platanenallee 6, D–15735 Zeuthen, Germany*

**Abstract**

The gluon contributions to $F_L(x, Q^2)$ in $\mathcal{O}(\alpha_s)$ are calculated taking into account the transverse momentum of the initial state parton. In comparison with collinear factorization $F_L(x, Q^2)$, is not affected at large $x$ but takes smaller values in the small $x$ range. The onset of the $k_\perp$ effect is a function of $Q^2$.

# Small $x$ Contributions to the Structure Function $F_L(x,Q^2)$

J. Blümlein[a]

[a]DESY–Zeuthen, Platanenallee 6, D–15735 Zeuthen, Germany

The gluon contributions to $F_L(x,Q^2)$ in $\mathcal{O}(\alpha_s)$ are calculated taking into account the transverse momentum of the initial state parton. In comparison with collinear factorization, $F_L(x,Q^2)$ is not affected at large $x$ but takes smaller values in the small $x$ range. The onset of the $k_\perp$ effect is a function of $Q^2$.

## 1. Introduction

In the small $x$ range a novel behaviour of nucleon structure functions is expected. Among possible dynamical effects are those due to non strong $k_\perp$ ordering [1] and screening [2]. Their description requires a generalization the factorization of the hadronic matrix elements. The $k_\perp$ dependence of the parton distributions can no longer be neglected in the hard scattering cross sections and $K^2 = -k_\perp^2$ dependent parton distributions must be used [3], i.e.

$$|\mathcal{M}|^2 \sim \int^{\mu^2} dK^2 \hat\sigma(x,K^2,\mu^2) \otimes \frac{\partial xG(x,K^2)}{\partial K^2} \qquad (1)$$

instead of the collinear relation

$$|\mathcal{M}|^2 \sim \hat\sigma(x,K^2=0,\mu^2) \otimes xG(x,\mu^2). \qquad (2)$$

The present paper aims on finding a consistent solution of the above problem without any approximations of the $x$ and $k_\perp$ behaviour of the coefficient functions. In this way former investigations [4,5] are extended. We aim on a *general formulation* of the gluon contribution to structure functions which offers the possibility to unfold the $k_\perp$ dependence of the gluon distribution at small $x$. Theoretical predictions of its small $x$ behaviour can thus be *directly* compared with the data and test the $k_\perp$ dependence.

## 2. $k_\perp$ Factorization

The $K^2$ integral in (1) extends to $K^2 = 0$. However, a perturbative definition of a gluon distribution is only possible at suitably large virtualities.

Therefore, we use [6]

$$\begin{aligned}
F_L^g(x,Q^2) &= \int_x^1 \frac{dz}{z} f_L^{g,0}(z) \frac{x}{z} G(z,Q_0^2) \\
&+ \int_x^1 \frac{dz}{z} \int_{Q_0^2}^{K_{max}^2} dK^2 f_L^g(z,\frac{K^2}{Q^2}) \frac{x}{z} \frac{\partial G(x/z,K^2)}{\partial K^2} \\
&\times \theta(K_{max}^2 - Q_0^2)
\end{aligned} \qquad (3)$$

where $K_{max}^2 = Q^2(1-z)/z$. We introduced a scale $Q_0$ for which we demand that $Q_0^2 \ll Q^2$. (3) is equivalent to (1) up to terms of $\mathcal{O}((Q_0^2/Q^2)^n)$. Note that eq. (3) contains the gluon distribution $G(x,K^2)$ only at virtualities in the perturbative range.

## 3. $F_L(x,Q^2)$ in $O(\alpha_s)$

For the gluonic contribution to $F_L(x,Q^2)$ eq. (3) the coefficient function takes the following form

$$\begin{aligned}
f_L^g(z,\frac{K^2}{Q^2}) &= \frac{2}{\pi}\alpha_s(Q^2) \sum_{q=1}^{N_f/2} (e_{q_u}^2 + e_{q_d}^2) \\
&\times \left\{ \frac{1}{64z} \left(\frac{Q^2}{K^2}\right)^2 G_{1L}^{(0,4)}(\omega,\beta) \right. \\
&+ z\frac{1}{16}\frac{Q^2}{K^2} \left[ G_{2L}^{(2,4)}(\omega,\beta) \right. \\
&+ \left.\left. G_{3L}^{(1,5)}(\omega,\beta) \log\left|\frac{1-\omega}{1+\omega}\right| \right] \right\}
\end{aligned} \qquad (4)$$

with $\omega = \sqrt{1 - 4K^2 z/Q^2}$, $\zeta = 4K^2 z/Q^2$, $\cos\beta = (1-\zeta/2)/\sqrt{1-z\zeta}$, and $G_{iL}^{(a,b)}(\omega,\beta) = \sum_{k=a}^{b} g_{ki}^{(L)}(\beta)/\omega^k$. The coefficients $g_{ki}$ are:

$$g_{01}^{(L)}(\beta) = \frac{5}{2} + 3\cos^2\beta - \frac{3}{2}\cos^4\beta$$

$$\begin{aligned}
g_{11}^{(L)}(\beta) &= 4\cos\beta - 12\cos^3\beta \\
g_{21}^{(L)}(\beta) &= 3 - 18\cos^2\beta + 15\cos^4\beta \\
g_{31}^{(L)}(\beta) &= -12\cos\beta + 20\cos^3\beta \\
g_{41}^{(L)}(\beta) &= -\frac{3}{2} + 15\cos^2\beta - \frac{35}{2}\cos^4\beta \\
g_{22}^{(L)}(\beta) &= 4 - 12\cos^2\beta \\
g_{23}^{(L)}(\beta) &= -24\cos\beta + 40\cos^3\beta \\
g_{24}^{(L)}(\beta) &= -3 + 30\cos^2\beta - 35\cos^4\beta \\
g_{31}^{(L)}(\beta) &= \frac{3}{4} + \frac{1}{2}\cos^2\beta + \frac{3}{4}\cos^4\beta \\
g_{32}^{(L)}(\beta) &= 2\cos\beta - 6\cos^3\beta \\
g_{33}^{(L)}(\beta) &= \frac{7}{2} - 15\cos^2\beta + \frac{15}{2}\cos^4\beta \\
g_{34}^{(L)}(\beta) &= -18\cos\beta + 30\cos^3\beta \\
g_{35}^{(L)}(\beta) &= -\frac{9}{4} + \frac{45}{2}\cos^2\beta - \frac{105}{4}\cos^4\beta \quad (5)
\end{aligned}$$

In the limit $K^2 \to 0$ one obtains the well-known result [9]

$$f_L^{g,0}(z) = \frac{2}{\pi}\alpha_s(Q^2) \sum_{q=1}^{N_f/2} \left(e_{q_u}^2 + e_{q_d}^2\right) z^2(1-z). \quad (6)$$

## 4. Numerical Results

Figure 1 shows the logarithmic derivative of the gluon distribution $dxG(x,Q^2)/d\log Q^2$ for different sets of parton parametrizations in the $\overline{\text{MS}}$ scheme. The most recent results, CTEQ2M and MRSA, were determined using the data measured at HERA, and do practically coincide, while earlier ones show some variation at small $x$. We will refer to the CTEQ2 parametrization [8] as an input in the following.

In figure 2 the gluonic contributions to $F_L(x,Q^2)$ using either eq. (6) or (3) are compared. At large $x$ coinciding results are obtained, but at small $x$ the collinear approach yields larger values for $F_L$. Setting $\mu^2 = Q^2(1-z)/z$, the kinematical upper limit of the $K^2$ integral (3), instead of $\mu^2 = Q^2$, in (6) leads to a lowering of $F_L(x,Q^2)$ already. Figure 2 shows that with rising $Q^2$, the effect due to finite $k_\perp$ emerges at smaller values of $x$. As expected, the onset of small $x$ effects is $Q^2$ dependent.

The separation scale $Q_0^2$ required in (3) affects $F_L(x,Q^2)$ very weakly as long as $Q^2 \gg Q_0^2$, which we assume. This is illustrated in figure 3. The effect of this choice of scale is comparable to that of $Q_0'^2$, the starting point of QCD evolution.

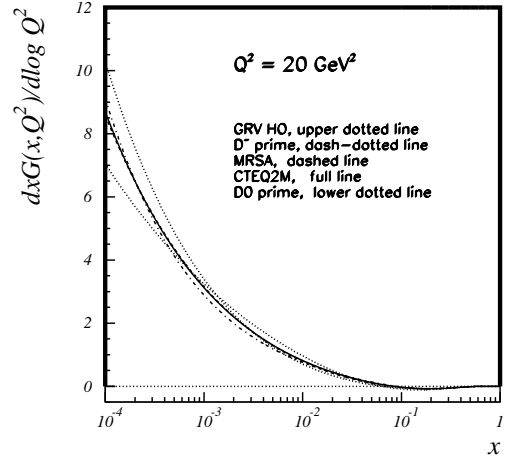

**Fig. 1** Logarithmic slope of the gluon momentum distribution vs $x$ for different parton parametrizations [7].

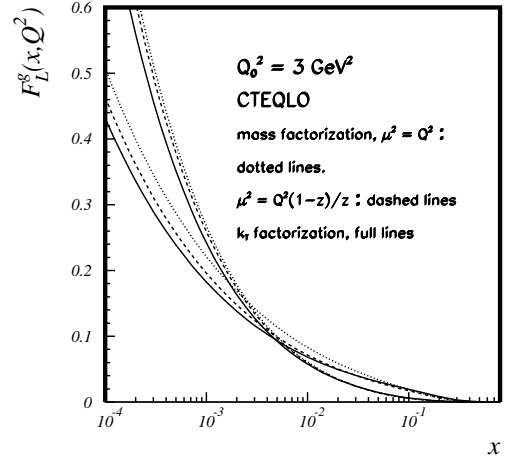

**Fig. 2** Comparison of the gluonic contributions to $F_L(x,Q^2)$ in the collinear case and $k_\perp$ factorization. The steeper lines are at $Q^2 = 10^4\,\text{GeV}^2$, the others at $Q^2 = 20\,\text{GeV}^2$.

In figure 4, the $\mathcal{O}(\alpha_s)$ result using $k_\perp$ factorization is compared with results of a $\mathcal{O}(\alpha_s^2)$ calcu-lation in the collinear approach [10]. The gluon contribution to $F_L$ in the collinear approach is diminished by about 10% by the $\mathcal{O}(\alpha_s^2)$ term for $x \sim 10^{-4}$ and $Q^2 \sim \mathcal{O}(20\,\text{GeV}^2)$. The $\mathcal{O}(\alpha_s)$ value of $F_L$ using $k_\perp$ factorization is somewhat smaller than the $\mathcal{O}(\alpha_s^2)$ value in the collinear approach. Note that the results are nearly equal in the range $x \sim 10^{-4}$. The quark contribution to $F_L$ in $\mathcal{O}(\alpha_s^2)$ [10] amounts to $\sim 10\%$ at $x \sim 10^{-4}$.

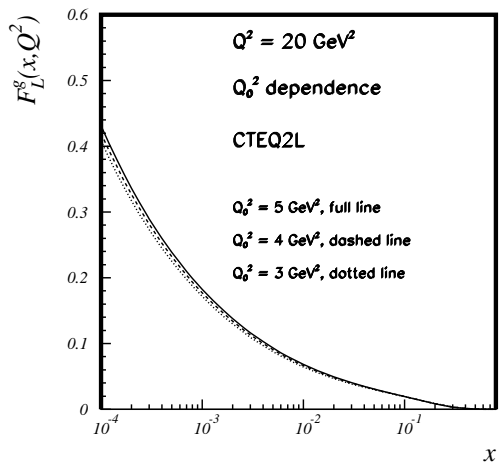

**Fig. 3** Dependence of $F_L^g(x, Q^2)$ on the choice of the separation scale $Q_0^2$.

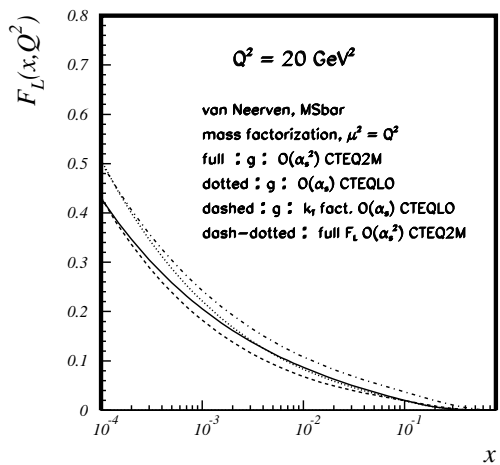

**Fig. 4** Comparison of the $\mathcal{O}(\alpha_s^2)$ calculation [10] with the result obtained in the $k_\perp$ factorization scheme.

## 5. Conclusions

A representation of $k_\perp$ factorization which is consistent with perturbative QCD has been given. The gluon contribution to the structure function $F_L(x, Q^2)$ was calculated using $k_\perp$ factorization *without* using any approximations of the Mellin convolution or the $x$ dependence of the coefficient functions, unlike some earlier investigations. The contributions to the structure functions obtained are *positive* in the whole $x$ range.

The derived coefficient functions approach those found using mass factorization in the limit $K^2 \to 0$. The numerical value obtained in $k_\perp$ factorization for suitably 'large' values of $x$ approach the result which ignores the $k_\perp$ dependence of the coefficient functions. This has been an expectation in the parton model [3]. There is *no fixed onset* (e.g. $x \sim 10^{-2}$ [5]) of the small $x$ effects observed. Deviations from the collinear result become smaller with rising $Q^2$ at constant $x$. The effect of the separation scale $Q_0$ is found to be subleading.

The $k_\perp$ dependence of the coefficient function and gluon distribution results into *smaller values* of $F_L$ in $\mathcal{O}(\alpha_s)$ in the small $x$ range. Quite similar values are obtained for $F_L$ in $\mathcal{O}(\alpha_s^2)$ [10] using mass factorization.

For discussions I would like to thank to W. van Neerven, J. Botts, and S. Catani.